\documentclass[fleqn,usenatbib]{mnras}

\usepackage{newtxtext,newtxmath}

\usepackage[T1]{fontenc}

\DeclareRobustCommand{\VAN}[3]{#2}
\let\VANthebibliography\thebibliography
\def\thebibliography{\DeclareRobustCommand{\VAN}[3]{##3}\VANthebibliography}

\usepackage{graphicx}	
\usepackage{amsmath}	
\usepackage{amssymb}	

\title[2020 AV$_2$ Orbital Dynamics]{Orbital Dynamics of 2020 AV$_2$: the First Vatira Asteroid}

\author[S. Greenstreet]{
Sarah Greenstreet$^{1,2}$\thanks{E-mail: sarah@b612foundation.org, sarahjg@uw.edu}
\\
$^{1}$Asteroid Institute, 20 Sunnyside Ave, Suite 427, Mill Valley, CA 94941\\
$^{2}$Department of Astronomy and the DIRAC Institute, University of Washington, 3910 15th Avenue NE, Seattle, WA 98195, USA\\
}

\date{Accepted 2020 February 5. Received 2020 February 5; in original form 2020 January 24}

\pubyear{2020}

\begin{document}
\label{firstpage}
\pagerange{\pageref{firstpage}--\pageref{lastpage}}
\maketitle

\begin{abstract}
Vatira-class near-Earth objects (NEOs) have orbits entirely interior to the orbit of Venus with aphelia $0.307<Q<0.718$~AU. Recently discovered asteroid 2020 AV$_2$ by the Zwicky Transient Facility on 4 January 2020 is the first known object on a Vatira orbit. Numerical integrations of 2020 AV$_2$'s nominal orbit show it remaining in the Vatira region for the next few hundred kyr before coupling to Venus and evolving onto an Atira orbit (NEOs entirely interior to Earth's orbit with $0.718<Q<0.983$~AU) and eventually scattering out to Earth-crossing. The numerical integrations of 9900 clones within 2020 AV$_2$'s orbital uncertainty region show examples of Vatira orbits trapped in the 3:2 mean-motion resonance with Venus at semimajor axis $a\approx0.552$~AU that can survive on the order of a few Myr. Possible 2020 AV$_2$ orbits also include those on Vatira orbits between Mercury and Venus that only rarely cross that of a planet. Together the 3:2 resonance and these rarely-planet-crossing orbits provide a meta-stable region of phase space that are stable on timescales of several Myr. If 2020 AV$_2$ is currently in this meta-stable region (or was in the past), that may explain its discovery as the first Vatira and may be where more are discovered.
\end{abstract}

\begin{keywords}
celestial mechanics -- minor planets, asteroids: individual: Vatiras -- minor planets, asteroids: individual: Atiras -- planets and satellites: dynamical evolution and stability
\end{keywords}



\section{Introduction}

The dynamical population of Vatira-class near-Earth objects (NEOs) was first discussed by \citet{Greenstreetetal2012} who defined them to be NEOs entirely interior to Venus' orbit with aphelia $0.307<Q<0.718$ AU, distinct from Atira-class NEOs with $0.718<Q<0.983$ AU that are decoupled from Earth's orbit. The recent discovery of 2020 AV$_2$ (semimajor axis $a=0.5556\pm0.0003$ AU, eccentricity $e=0.1767\pm0.0006$, inclination $i=15.86^o\pm0.02^o$, longitude of ascending node $\Omega=6.702^o\pm0.007^o$, and argument of perihelion $\omega=187.27^o\pm0.05^o$ from JPL's Small-Body Database\footnote{http://ssd.jpl.nasa.gov/sbdb.cgi} \citep{Giorgini2015}) by the Zwicky Transient Facility on 4 January 2020 \citep{Baccietal2020} marks the first known Vatira asteroid. With $H=16.4$, 2020 AV$_2$ is likely one of two Vatiras in existence of this size, given there are $\sim$1,000 $H<18$ NEOs of which $\simeq0.22\%$ are Vatiras \citep{Greenstreetetal2012, Granviketal2018}.

NEOs reach the Vatira region through Earth, Venus, and Mercury close encounters. The NEO ``source mapper" available with the \citet{Greenstreetetal2012} NEO orbital distribution model release\footnote{Downloadable at http://www.sarahgreenstreet.com/neo-model.} shows that 2020 AV$_2$ most likely originated from the $\nu_6$ secular resonance (the mean probability that it came from the $\nu_6$ is $0.70\pm0.21$, the 3:1 mean-motion resonance with Jupiter is $0.14\pm0.17$, and the intermediate Mars crossing population is $0.16\pm0.16$). Typical Vatiras are expected to have $e\simeq0.4$ and $i\simeq25^o$, although 2020 AV$_2$ has smaller $e$ and $i$ than the average Vatira.

\section{Orbital Stability of 2020 AV$_2$}

\begin{figure}
	\includegraphics[width=\columnwidth]{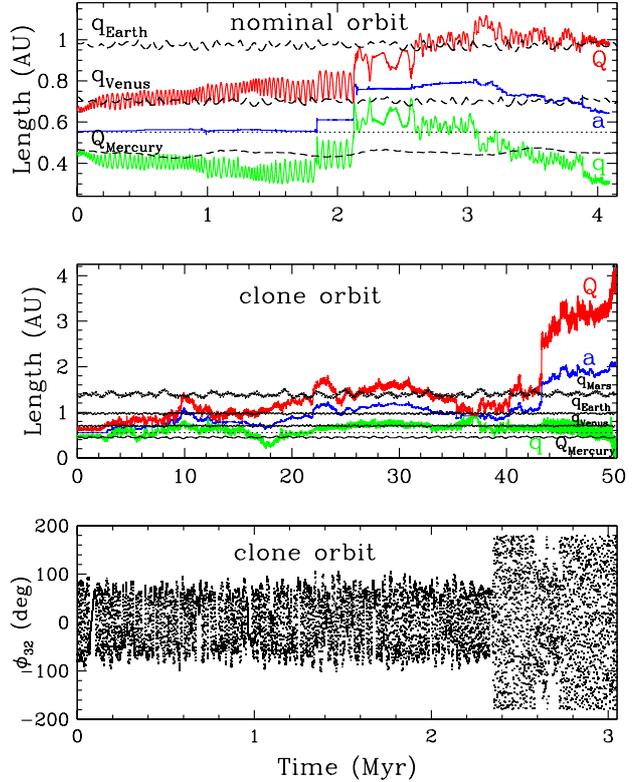}
    \caption{Aphelion (red), semimajor axis (blue), and perihelion (green) versus time for the nominal orbit of 2020 AV$_2$ (top) and a clone within 2020 AV$_2$'s orbital uncertainty region (middle) as well as the venusian 3:2 resonant argument $\phi_{32}$ for the clone (bottom). Note the different length and time scales in the three panels. The dotted line at $\approx0.552$ AU shows the location of the 3:2 mean-motion resonance with Venus. The dashed lines show $Q_{\rm Mercury}$, $q_{\rm Venus}$, $q_{\rm Earth}$, and $q_{\rm Mars}$. Libration of $\phi_{32}$ around $0^o$ indicates the clone orbit is in the 3:2 resonance for the next $\approx2.4$~Myr and again briefly from $\approx2.6-2.7$~Myr from now. After this time, the clone scatters away from the resonant semimajor axis.}
    \label{Fig:Q_a_q_phi32}
\end{figure}

Vatira orbits are predicted to be commonly unstable due to repeated Venus and Mercury encounters, which often cause oscillation between the Vatira and Atira states. Many Vatira's can experience Kozai oscillations in $e$ and $i$ \citep{Kozai1962,MichelThomas1996}, which aid the Vatira/Atira state oscillation. Numerical integrations using the N-body code SWIFT-RMVS4 with a $\approx4$~hr default time step and the inclusion of the planets Mercury-Saturn and point-mass Newtonian gravitational effects only were run to determine the stability of 2020 AV$_2$'s nominal orbit. The nominal orbit does not remain decoupled from Venus (Figure~\ref{Fig:Q_a_q_phi32}'s top panel), similar to typical Vatiras (see figures 7 \& 8 of \citet{Greenstreetetal2012}). $\approx0.14$~Myr from now, $Q>q_{\rm Venus}$ and 2020 AV$_2$ then oscillates between the Vatira and Atira states until $\approx1.20$~Myr from now at which time it leaves the Vatira state. Up to this time, 2020 AV$_2$ remains mostly Mercury-crossing, but at $\approx1.84$~Myr from now, $q$ begins oscillating around $Q_{\rm Mercury}$ before fully decoupling from Mercury at $\approx2.11$~Myr. $\approx0.48$~Myr later at $\approx2.59$~Myr from now, the object begins oscillating between the Atira and Aten ($Q>0.983$~AU, $a<1.0$~AU) states where it remains. At $\approx3.33$~Myr, it re-couples to Mercury and then collides with Venus (the most common Vatira end-state \citep{Greenstreetetal2012}) $\approx4.1$~Myr after the start of the integration. Integrated over the next 4.1~Myr, 2020 AV$_2$ spends $\approx0.44$~Myr in the Vatira region, roughly twice that of most Vatiras ($\sim0.25$~Myr integrated over their lifetimes \citep{Greenstreetetal2012}).

The nominal orbit of 2020 AV$_2$ is quite uncertain due its 12 day arc length. To better understand the current stability of 2020 AV$_2$ within the Vatira region, 9900 clones within the orbital uncertainty region from JPL's Scout NEO Hazard Assessment page\footnote{https://cneos.jpl.nasa.gov/scout/intro.html} on 7 Jan 2020 were also integrated for 2~Myr (using the same integration parameters as described above). By this time, $>99.7\%$ of the clones were still active; six collided with Mercury, 22 with Venus. $\approx29\%$ of the total integrated clone residence time is in the Vatira region; on average, clones spend $\approx0.58$~Myr in the Vatira region integrated over their lifetimes, which is slightly longer than the nominal orbit. The median time at which the clones re-couple to Venus is $\approx0.14$~Myr from now, similar to the nominal orbit.

While 2020 AV$_2$'s nominal orbit only rarely decouples from (is entirely exterior to) Mercury while a Vatira, Figure~\ref{Fig:Q_a_q_phi32}'s middle panel shows a 2020 AV$_2$ clone that spends the vast majority of the next $\approx2.4$~Myr on an orbit between Mercury and Venus that only rarely crosses Mercury's orbit. This clone also has $a\approx0.552$~AU, which is the location of the 3:2 mean-motion resonance with Venus. Figure~\ref{Fig:Q_a_q_phi32}'s bottom panel shows the 3:2 resonant argument $\phi_{32}$ librating around $0^o$ indicating this clone is within the resonance for the next $\approx2.4$~Myr (note the different time scales between the three panels in Figure~\ref{Fig:Q_a_q_phi32}).

At the time it leaves the resonance, the clone is still in the Vatira region and becomes more strongly coupled to Mercury, which likely kicked it out of the resonance. $\approx0.1$~Myr later, the clone couples to Venus and evolves onto an Atira orbit. It then briefly re-enters the 3:2 resonance at $\approx2.6$~Myr from now for another $\approx0.1$~Myr where it is protected from Venus close encounters before once again leaving the resonance (again, likely due to a Mercury close encounter) and fully re-coupling to Venus as an Atira. It then remains an Atira for $\approx5.8$~Myr before first coupling to Earth $\approx8.2$~Myr from now. The clone lives for another $\approx40$~Myr, during which time the clone visits the Atira, Aten, and Apollo ($a>1.0$~AU, $q<1.017$~AU) NEO states multiple times while random walking in semimajor axis due to planetary close encounters with Mercury, Venus, Earth, and eventually Mars. The clone is pushed into the Sun ($q<0.005$~AU; the second most common end-state for Vatiras \citep{Greenstreetetal2012}) $\approx50.2$~Myr after the start of the integration.

Although the 3:2 resonance protects this clone from close encounters with Venus during the first $\approx2.4$~Myr of the integration, its orbit is not Venus-crossing for the vast majority of the time it is in the resonance, and since it only rarely crosses Mercury's orbit, the resonance only adds additional stability on top of its already meta-stable, rarely-planet-crossing orbit. Although nearly all clones spend time near $a_{32}\approx0.552$~AU, it is for $<1\%$ of the total integrated clone residence time. In addition, the clones only spend $\approx3.5\%$ of their total integrated residence time entirely exterior to Mercury's orbit. However, this clone is evidence that there are 2020 AV$_2$ orbits within the uncertainty region that are in a meta-stable portion of phase space where Vatiras can live on the order of a few Myr (an order of magnitude longer than typical Vatiras can remain in the Vatira region integrated over their lifetimes) on non- or rarely-planet-crossing orbits in or near the venusian 3:2 mean-motion resonance.

In a dynamical study of the Atira population, \citet{Ribeiroetal2016} show stable regions of phase space between Mercury and Venus ($a\approx0.5-0.65$~AU) at low-eccentricity ($e\lesssim0.2$) and low-inclination ($i<30^0$) in the Vatira region, but do not discuss this stability in the context of the Vatira population; the orbital uncertainty region of 2020 AV$_2$ falls within this stable region. Interestingly, their figure 2 shows that the location of the 3:2 mean-motion resonance with Venus is less stable than the regions just outside the resonance. 

After submission of this paper, the author became aware of a recently submitted paper by \citet{dlFM2_2020} on the orbital evolution of 2020 AV$_2$ that reports similar findings to those discussed here. Their paper shows evidence of a dynamical pathway for Vatiras, including 2020 AV$_2$, to enter the 3:2 mean-motion resonance with Venus and report that a possible long-term stable population of Vatiras may be trapped in the 3:2 resonance while undergoing von Zeipel-Lidov-Kozai oscillations in $e$ and $i$, similar to the discussion above.

\section{Conclusions}

2020 AV$_2$ likely originated in the $\nu_6$ secular resonance in the main belt before evolving onto an Earth-crossing orbit where subsequent Earth, Venus, and Mercury close encounters scattered it down to its current location between Mercury and Venus. Its nominal orbit is that of a currently relatively long-lived Vatira where it will likely remain for a few hundred kyr before re-coupling to Venus and scattering back out through the Atira state to the Earth-crossing region, eventually most likely colliding with Venus. Although 2020 AV$_2$'s nominal orbit is not currently in the 3:2 mean-motion resonance with Venus, it may very well have been in the past, where it may have resided for several Myr. It may also currently reside on a meta-stable, rarely-planet-crossing orbit in or near the 3:2 resonance where it could remain on the order of a few Myr. If 2020 AV$_2$ is currently in this meta-stable (near) resonant region (or was in the past), that may explain its discovery as the first Vatira and may be where more are discovered.

\section*{Acknowledgements}

S. Greenstreet would like to thank Brett Gladman and Mario Juri\'{c} for insightful comments, C. de la Fuente Marcos \& R. de la Fuente Marcos for comments on their own results on the orbital evolution of 2020 AV$_2$, and Lynne Jones for encouragement to start this project.

S. Greenstreet acknowledges support from the Asteroid Institute, a program of B612, 20 Sunnyside Ave, Suite 427, Mill Valley, CA 94941. Major funding for the Asteroid Institute was generously provided by the W.K. Bowes Jr. Foundation and Steve Jurvetson. Research support is also provided from Founding and Asteroid Circle members K. Algeri-Wong, B. Anders, R. Armstrong, G. Baehr, The Barringer Crater Company, B. Burton, D. Carlson, S. Cerf, V. Cerf, Y. Chapman, J. Chervenak, D. Corrigan, E. Corrigan, A. Denton, E. Dyson, A. Eustace, S. Galitsky, L. \& A. Fritz, E. Gillum, L. Girand, Glaser Progress Foundation, D. Glasgow, A. Gleckler, J. Grimm, S. Grimm, G. Gruener, V. K. Hsu \& Sons Foundation Ltd., J. Huang, J. D. Jameson, J. Jameson, M. Jonsson Family Foundation, D. Kaiser, K. Kelley, S. Krausz, V. La\v{s}as, J. Leszczenski, D. Liddle, S. Mak, G.McAdoo, S. McGregor, J. Mercer, M. Mullenweg, D. Murphy, P. Norvig, S. Pishevar, R. Quindlen, N. Ramsey, P. Rawls Family Fund, R. Rothrock, E. Sahakian, R. Schweickart, A. Slater, Tito's Handmade Vodka, T. Trueman, F. B. Vaughn, R. C. Vaughn, B. Wheeler, Y. Wong, M. Wyndowe, and nine anonymous donors.

S. Greenstreet acknowledges the support from the University of Washington College of Arts and Sciences, Department of Astronomy, and the DIRAC Institute. The DIRAC Institute is supported through generous gifts from the Charles and Lisa Simonyi Fund for Arts and Sciences and the Washington Research Foundation.




\bibliographystyle{mnras}
\bibliography{bibliography} 

\begin{thebibliography}{}
\makeatletter
\relax
\def\mn@urlcharsother{\let\do\@makeother \do\$\do\&\do\#\do\^\do\_\do\%\do\~}
\def\mn@doi{\begingroup\mn@urlcharsother \@ifnextchar [ {\mn@doi@}
  {\mn@doi@[]}}
\def\mn@doi@[#1]#2{\def\@tempa{#1}\ifx\@tempa\@empty \href
  {http://dx.doi.org/#2} {doi:#2}\else \href {http://dx.doi.org/#2} {#1}\fi
  \endgroup}
\def\mn@eprint#1#2{\mn@eprint@#1:#2::\@nil}
\def\mn@eprint@arXiv#1{\href {http://arxiv.org/abs/#1} {{\tt arXiv:#1}}}
\def\mn@eprint@dblp#1{\href {http://dblp.uni-trier.de/rec/bibtex/#1.xml}
  {dblp:#1}}
\def\mn@eprint@#1:#2:#3:#4\@nil{\def\@tempa {#1}\def\@tempb {#2}\def\@tempc
  {#3}\ifx \@tempc \@empty \let \@tempc \@tempb \let \@tempb \@tempa \fi \ifx
  \@tempb \@empty \def\@tempb {arXiv}\fi \@ifundefined
  {mn@eprint@\@tempb}{\@tempb:\@tempc}{\expandafter \expandafter \csname
  mn@eprint@\@tempb\endcsname \expandafter{\@tempc}}}

\bibitem[\protect\citeauthoryear{{Bacci} et~al.,}{{Bacci}
  et~al.}{2020}]{Baccietal2020}
{Bacci} P.,  et~al., 2020, Minor Planet Electronic Circular, \href
  {https://ui.adsabs.harvard.edu/abs/2020MPEC....A...99B/abstract} {pp
  2020--A99}

\bibitem[\protect\citeauthoryear{{Giorgini}}{{Giorgini}}{2015}]{Giorgini2015}
{Giorgini} J.~D.,  2015, IAU General Assembly, \href
  {https://ui.adsabs.harvard.edu/abs/2015IAUGA..2256293G/abstract} {29,
  2256293}

\bibitem[\protect\citeauthoryear{{Granvik} et~al.,}{{Granvik}
  et~al.}{2018}]{Granviketal2018}
{Granvik} M.,  et~al., 2018, \mn@doi [Icarus] {10.1016/j.icarus.2018.04.018},
  \href {https://ui.adsabs.harvard.edu/abs/2018Icar..312..181G/abstract} {312,
  181}

\bibitem[\protect\citeauthoryear{{Greenstreet}, {Gladman}  \&
  {Ngo}}{{Greenstreet} et~al.}{2012}]{Greenstreetetal2012}
{Greenstreet} S.,  {Gladman} B.,   {Ngo} H.,  2012, Icarus, \href
  {https://ui.adsabs.harvard.edu/abs/2012Icar..217..355G/abstract} {217, 355}

\bibitem[\protect\citeauthoryear{{Kozai}}{{Kozai}}{1962}]{Kozai1962}
{Kozai} Y.,  1962, \mn@doi [The Astronomical Journal] {10.1086/108790}, \href
  {https://ui.adsabs.harvard.edu/abs/1962AJ.....67..591K/abstract} {67, 591}

\bibitem[\protect\citeauthoryear{{Michel} \& {Thomas}}{{Michel} \&
  {Thomas}}{1996}]{MichelThomas1996}
{Michel} P.,  {Thomas} F.,  1996, \mn@doi [Astronomy \& Astrophysics]
  {1996A&A...307..310M}, \href
  {https://ui.adsabs.harvard.edu/abs/1996A%26A...307..310M/abstract} {786, 310}

\bibitem[\protect\citeauthoryear{{Ribeiro}, {Roig}, {De~Pr\'{a}}, {Carvano}  \&
  {DeSouza}}{{Ribeiro} et~al.}{2016}]{Ribeiroetal2016}
{Ribeiro} A.~O.,  {Roig} F.,  {De~Pr\'{a}} M.~N.,  {Carvano} J.~M.,   {DeSouza}
  S.~R.,  2016, \mn@doi [Monthly Notices of the Royal Astronomical Society]
  {10.1093/mnras/stw642}, \href
  {https://ui.adsabs.harvard.edu/abs/2016MNRAS.458.4471R/abstract} {458, 4471}

\bibitem[\protect\citeauthoryear{{de la Fuente Marcos} \& {de la Fuente
  Marcos}}{{de la Fuente Marcos} \& {de la Fuente Marcos}}{2020}]{dlFM2_2020}
{de la Fuente Marcos} C.,  {de la Fuente Marcos} R.,  2020, Monthly Notices of
  the Royal Astronomical Society, p. submitted

\makeatother
\end{thebibliography}








\bsp	
\label{lastpage}
\end{document}